\documentclass[
reprint,
superscriptaddress,
nofootinbib,
 amsmath,amssymb,
 aps,
prb,
]{revtex4-2}

\usepackage{graphicx}
\usepackage{dcolumn}
\usepackage{bm}
\usepackage[colorlinks,linkcolor=blue,anchorcolor=red,citecolor=blue,urlcolor=blue]{hyperref}

\usepackage{tabularx}
\usepackage{multirow} 
\usepackage{array} 
\usepackage{threeparttable}
\usepackage{booktabs}

\begin{document}


\title{Selective observation of surface and bulk bands in polar WTe$_2$ by laser-based spin- and angle-resolved photoemission spectroscopy}

\author{Yuxuan Wan}
\affiliation{Institute for Solid State Physics, the University of Tokyo, Kashiwa, Chiba 277-8581, Japan}
\affiliation{Department of Physics, the University of Tokyo, Bunkyo-ku, Tokyo 113-0033, Japan}

\author{Lihai Wang}
\affiliation{Laboratory for Pohang Emergent Materials, Pohang Accelerator Laboratory and Max Plank POSTECH Center for Complex Phase Materials, Pohang University of Science and Technology, Pohang 790-784, Korea}

\author{Kenta Kuroda}
\affiliation{Institute for Solid State Physics, the University of Tokyo, Kashiwa, Chiba 277-8581, Japan}

\author{Peng Zhang}
\affiliation{Institute for Solid State Physics, the University of Tokyo, Kashiwa, Chiba 277-8581, Japan}

\author{Keisuke Koshiishi}
\affiliation{Department of Physics, the University of Tokyo, Bunkyo-ku, Tokyo 113-0033, Japan}

\author{Masahiro Suzuki}
\affiliation{Department of Physics, the University of Tokyo, Bunkyo-ku, Tokyo 113-0033, Japan}

\author{Jaewook Kim}
\affiliation{Rutgers Center for Emergent Materials and Department of Physics and Astronomy, 136 Frelinghuysen Road, Piscataway, NJ 08854, USA}

\author{Ryo Noguchi}
\affiliation{Institute for Solid State Physics, the University of Tokyo, Kashiwa, Chiba 277-8581, Japan}

\author{C\'edric Bareille}
\affiliation{Institute for Solid State Physics, the University of Tokyo, Kashiwa, Chiba 277-8581, Japan}

\author{Koichiro Yaji}
\affiliation{Research Center for Advanced Measurement and Characterization, National Institute for Materials Science, 3-13, Sakura, Tsukuba, Ibaraki 305-0003, Japan}

\author{Ayumi Harasawa}
\affiliation{Institute for Solid State Physics, the University of Tokyo, Kashiwa, Chiba 277-8581, Japan}

\author{Shik Shin}
\affiliation{Institute for Solid State Physics, the University of Tokyo, Kashiwa, Chiba 277-8581, Japan}
\affiliation{Office of University Professor, The University of Tokyo, Chiba 277-8581, Japan}

\author{Sang-Wook Cheong}
\affiliation{Laboratory for Pohang Emergent Materials, Pohang Accelerator Laboratory and Max Plank POSTECH Center for Complex Phase Materials, Pohang University of Science and Technology, Pohang 790-784, Korea}
\affiliation{Rutgers Center for Emergent Materials and Department of Physics and Astronomy, 136 Frelinghuysen Road, Piscataway, NJ 08854, USA}

\author{Atsushi Fujimori}
\affiliation{Department of Physics, the University of Tokyo, Bunkyo-ku, Tokyo 113-0033, Japan}

\author{Takeshi Kondo}
\altaffiliation{Corresponding author: kondo1215@issp.u-tokyo.ac.jp}
\affiliation{Institute for Solid State Physics, the University of Tokyo, Kashiwa, Chiba 277-8581, Japan}
\affiliation{Trans-scale Quantum Science Institute, The University of Tokyo, Tokyo 113-0033, Japan}

\date{\today}

\begin{abstract}
The electronic state of WTe$_2$, a candidate of type-II Weyl semimetal, is investigated by using laser-based spin- and angle-resolved photoemission spectroscopy (SARPES). We prepare the pair of  WTe$_2$ samples, one with (001) surface and the other with ($00\overline{1}$) surface, by ``sandwich method”, and measure the band structures of each surface separately. The Fermi arcs are observed on both surfaces. We identify that the Fermi arcs on the two surfaces are both originating from surface states. We further find a surface resonance band, which connects with the Fermi-arc band, forming a Dirac-cone-like band dispersion. Our results indicate that the bulk electron and hole bands are much closer in momentum space than band calculations. 
\end{abstract}

\maketitle

Weyl semimetals host electronic quasiparticles behaving as Weyl fermions, which are massless chiral fermions derived from the Weyl equation \cite{weyl1929elektron}, and these have attracted much attention in recent years because of their promising novel physical properties \cite{hosur2012charge,Weng2015}. While Weyl fermions have not been observed in the field of fundamental particle physics, Weyl semimetals have been discovered experimentally \cite{wan2011topological,Xu2015,Xu2015a}: these demonstrated to have gapless cone-like linear dispersions around the topologically protected Weyl points, which are split from Dirac points due to either broken time-reversal or space-inversion symmetry. On the crystal surface, the topological surface states emerge, and intriguingly, these connect a pair of Weyl points that have opposite chiralities, forming arc-like Fermi surfaces (Fermi arcs) \cite{xu2016spin}. 

\begin{figure}
    \includegraphics[width=3in]{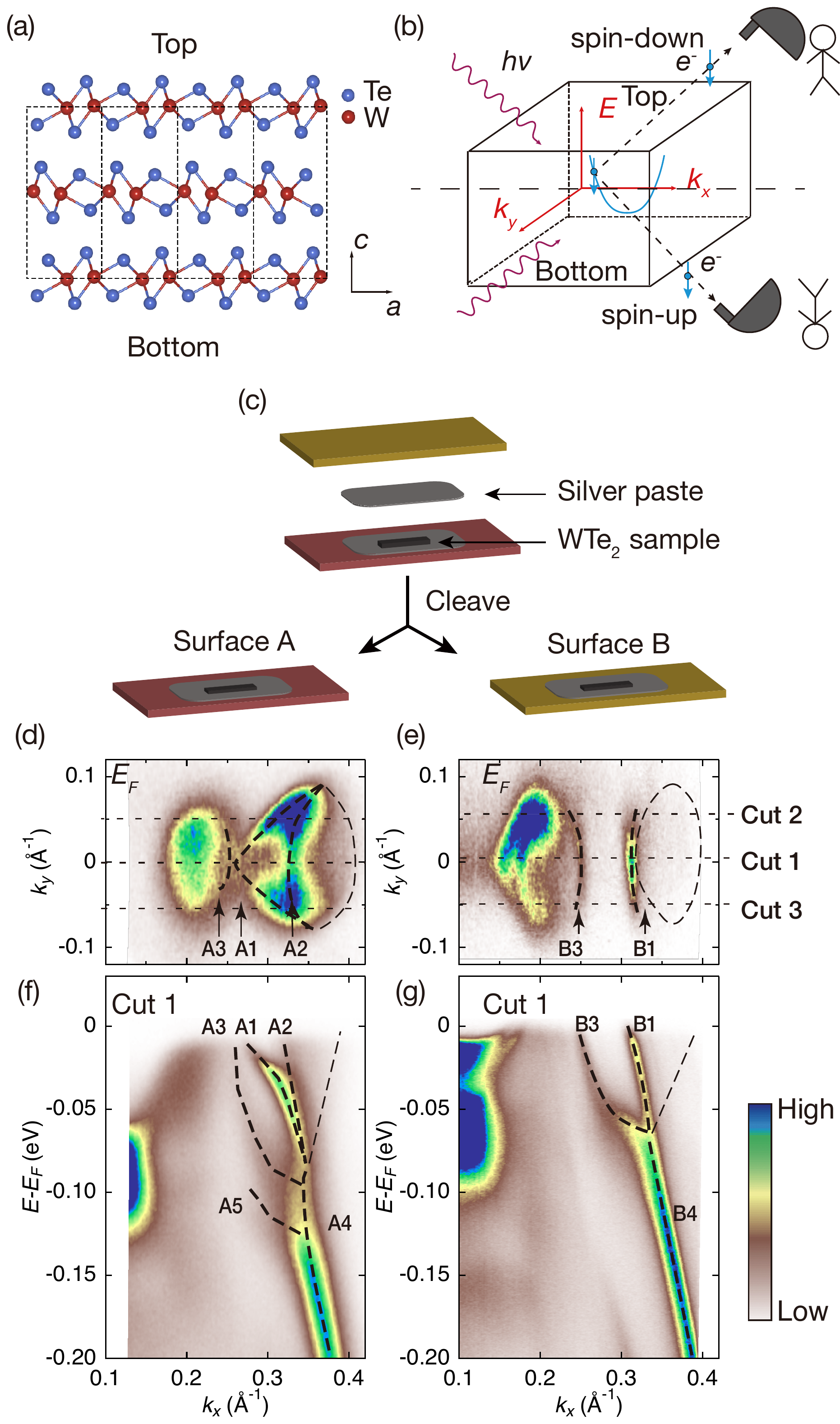}
    \caption{(a) The crystal structure of WTe$_2$ with different top and bottom surfaces. (b) Schematic of the SARPES measurements from opposite crystal surfaces; the spin direction of the bulk band becomes opposite when observing from opposite surfaces. (c) The ``sandwich method" we employed to get a pair of samples with surfaces A and B. (d)-(e) Fermi surfaces on surfaces A and B measured by laser-ARPES, respectively. (f)-(g) Band dispersions along the high symmetry cut 1 in (d) and (e), respectively. Here, the surface A would be corresponding to the surface of ``type B'' in Ref. \cite{Bruno2016} and ``type T'' in Ref. \cite{Wu2016}, and the surface B is to ``type A'' in Ref. \cite{Bruno2016} and ``type N'' in Ref. \cite{Wu2016}. 
The thick lines indicate the bands clearly seen in the data, whereas the thin dashed lines draw the bulk electron bands not clear in our data due to the matrix element effects.}
    \label{fig:Fig1}
\end{figure}

WTe$_2$ has a broken space-inversion symmetry with the polarization  
along the $c$-axis direction, and it is predicted to be a type-II Weyl semimetal \cite{Soluyanov2015}, where the tilted hole- and electron-like bands touch with each other at several Weyl points. Based on its unique electronic structures, WTe$_2$ has many novel properties, including giant non-saturating magnetoresistance \cite{Ali2014,Pletikosic2014}, superconductivity under high pressure \cite{kang2015superconductivity,pan2015pressure} or with potassium-intercalation \cite{Zhu2018a}, anomalous Hall effect \cite{zhang2018electrically,zhao2020observation}, and detection of the orbital angular momentum of light \cite{Ji2020}.
Weyl points of WTe$_2$ are expected to locate above the Fermi level. Thus, the direct observation of these  
by angle-resolved photoemission spectroscopy (ARPES) is not easy. Instead, many attempts have been made to identify the topological Fermi arc \cite{Pletikosic2014,Wu2015,Bruno2016,Wu2016,Wang2016,Wu2017,DiSante2017,Das2016,Feng2016}, which could be the hallmark of the Weyl semimetal state. Although WTe$_2$ has a polar structure with distinct (001) and ($00\overline{1}$) surfaces \cite{Zhou2016,Zhang2017,Sakano2017}[see Fig. \ref{fig:Fig1}(a)], most studies have been so far focused on one surface (surface A) which exhibits a large Fermi arc connecting bulk electron and hole pockets. In contrast, the detailed study on the other surface (surface B) is still lacking. Although the Fermi surface without a large Fermi arc has been also observed at different locations on a cleavage surface by ARPES, its origin is controversial. While such signals could be for the domain of surface B \cite{Bruno2016}, it may also be a result of the strain effect on the crystal surface which prevents the formation of the Fermi arc due to the transition from topological to trivial semimetal \cite{Wu2016}. To unravel the nature of the electronic structure in a type-II Weyl semimetal candidate WTe$_2$, one needs to solve this controversial issue and furthermore disentangle the relationship between the bulk and surface states, which is rather complex since both are spin-polarized due to the polar crystal structure with broken inversion symmetry.
 
To address these, in this paper, we employ new approaches in ARPES measurements. We prepare the (001) and ($00\overline{1}$) surfaces of WTe$_2$ from a single piece of crystal with the ``sandwich method", and separately observe the band structure for each surface by using laser-ARPES. This unique experiment demonstrates that Fermi arcs exist on both crystal surfaces, although each has different length. Furthermore, we use a laser-based spin- and angle-resolved photoemission spectroscopy (SARPES) to distinguish the surface and bulk states by comparing the spin-polarized spectra between the two surfaces, and verify that the Fermi arcs on both surfaces are surface states. We also find a surface resonance state derived from the bulk hole band, which connects to the Fermi arc band derived from the bulk electron band. The results indicate that the hole and electron bands are unexpectedly close to each other, causing the overlapping of the surface projection of these bands; consequently, the surface states of WTe$_2$ are intricately intertwined. 
 
High-quality single crystals of WTe$_2$ were grown by the iodine vapor transport method \cite{Zhang2017}. Laser-ARPES and SARPES measurements were performed with a 7 eV laser and a Scienta Omicron DA30-L analyzer equipped with twin very-low-energy-electron-diffraction (VLEED) spin-detectors in the Institute for Solid State Physics, the University of Tokyo \cite{Yaji2016,WTe2PRBSM}. The analyzer can use the deflector mode, which allows one to perform the momentum space ($k$-space) mapping without rotating the samples. The energy resolution was set to $\sim$3 meV and 20 meV in laser-ARPES and SARPES measurements, respectively. The spot size of the laser on the samples was $\sim$50 $\mu$m. 

The WTe$_2$ crystal has distinct top and bottom surfaces with different average separations between W and Te planes [Fig. \ref{fig:Fig1}(a)].  
To measure these opposite crystal surfaces separately, we employed a ``sandwich method", as shown in Fig. \ref{fig:Fig1}(c): 
both sides of a single crystal are glued on metal substrates, and then the crystal is split apart into two pieces. In this way, we can prepare two pieces of samples exposing opposite surfaces of a crystal (surfaces A and B). 
For such a pair of samples, we have separately performed laser-ARPES measurements at 30 K after cleaving \textit{in situ} at room temperature. 
On each surface, the band structure was the same at the different measurement positions, indicating that our samples have a large single domain.

\begin{figure*}
    \includegraphics[width=5in]{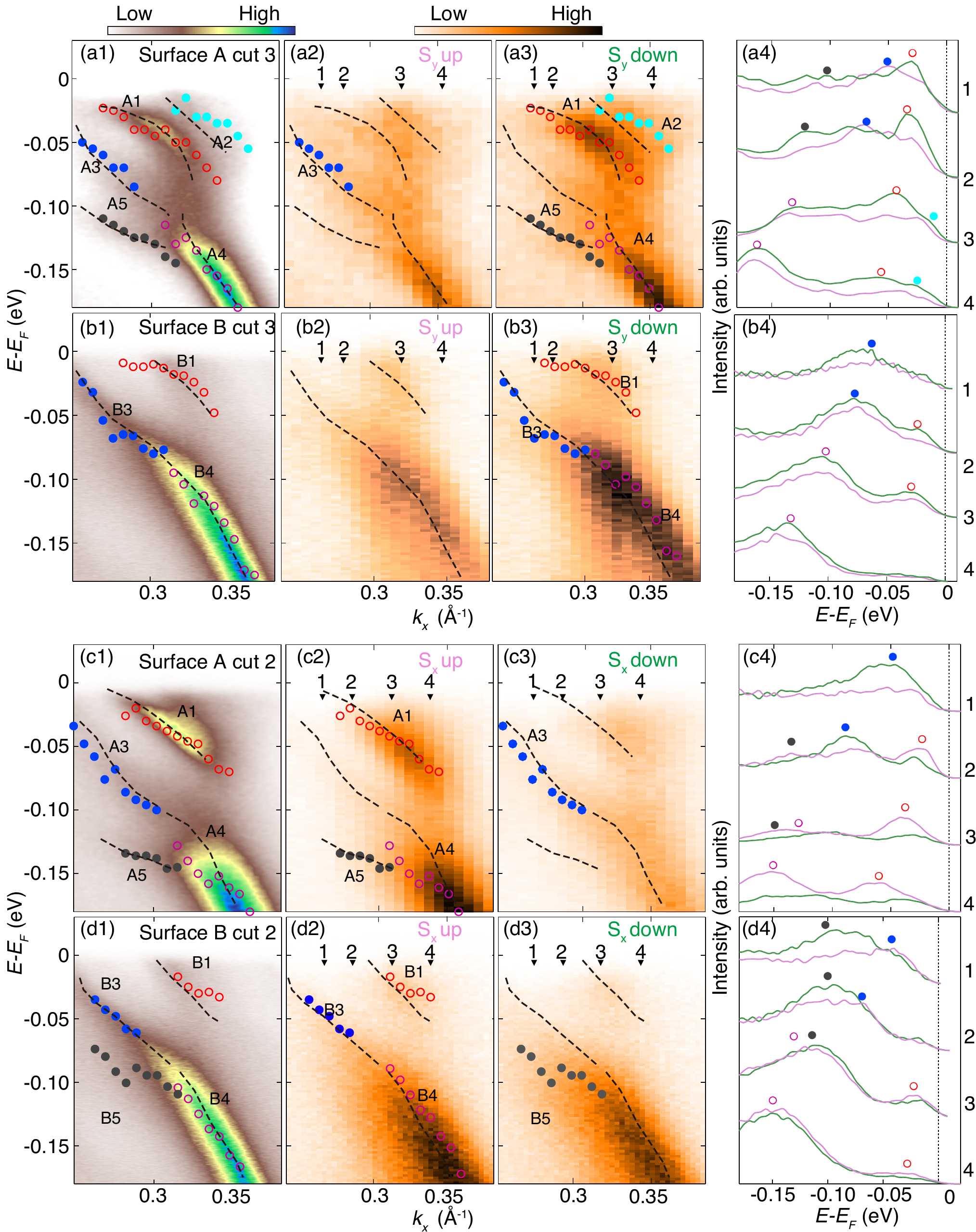}
    \caption{(a1)-(b1) The spin-integrated ARPES images along cut 3 [see Figs. \ref{fig:Fig1}(d) and (e)] on surface A and B, respectively. The dashed lines indicate the band dispersions guided from the spin-integrated ARPES images in (a1) and (b1). Circles are duplicated from SARPES results in (a2,a3) and (b2,b3). The corresponding bands on the opposite surface that are compared to each other share the same number. (a2)-(a3) SARPES images for $s_y$-up and -down, respectively, measured along cut 3 on surface A. (b2)-(b3) The same data as (a2)-(a3), but for surface B. The colored filled/open circles indicate the peaks of energy distribution curves (EDCs) for each bulk/surface state. The dashed lines are duplicated from spin-integrated ARPES results of (a1) and (b1). (a4) and (b4) EDCs for spin up (pink) and down (green) at the momenta marked by the triangles in (a2,a3) and (b2,b3), respectively. The circles mark the peak positions of dominant spectra. The colors of the plots are same as those of the corresponding bands. (c)-(d) The similar set of data to those in (a) and (b), but for spin-$x$ component observed along cut 2 [indicated in Figs. \ref{fig:Fig1}(d) and 1(e)] on surfaces A and B, respectively.}
    \label{fig:Fig2}
\end{figure*}

We have obtained quite different band structures (Fermi surface and energy dispersion) from surfaces A and B, as shown in Figs. \ref{fig:Fig1}(d,f) and Figs. \ref{fig:Fig1}(e,g), respectively. Here, $p$ polarized light was used. Surface A exhibits a large Fermi arc (A1)[Fig. \ref{fig:Fig1}(d)] connecting bulk electron and hole pockets [A2 and A3 in Figs. \ref{fig:Fig1}(d,f), respectively], which has been reproduced by band calculations and recognized as surface states \cite{Bruno2016,Wang2016,Wu2016}.  In contrast, while an arc-like feature (B1) is also detected on surface B [Fig. \ref{fig:Fig1}(e)], it is very short and cannot be reproduced by calculations. In the previous studies, these two kinds of band structures were obtained from different locations on one cleaved surface, and the reason for it was not clear yet \cite{Bruno2016,Wu2016}. Our experiments, shown in Fig. \ref{fig:Fig1}, clarify that the different band structures actually belong to opposite crystal surfaces [top and bottom surfaces in Fig. \ref{fig:Fig1}(a)]. Nevertheless, it is still controversial whether the short arc-like intensities observed on surface B originate from the surface state \cite{Bruno2016} or the bulk state as a portion of the electron pocket \cite{Wu2016}; thus, this issue needs to be resolved to understand the electronic properties of WTe$_2$.

In this study, SARPES is used to distinguish the surface states from the bulk states; the spin-polarization of all the bulk bands should be inverted when observed from the opposite surfaces (surface A and B), as depicted in Fig. \ref{fig:Fig1}(b). Hence, the bands showing the same spin-polarization in measurements can be, instead, defined as surface states \cite{Sakano2017}. As in Figs. \ref{fig:Fig1}(d)-(g), segments of the bands for surfaces A and B are labeled with letters and numbers, along with dashed lines overlaid on the ARPES dispersion. Here, the numbering such as in A4 and B4 are given to the corresponding bands for surface A and B which will be compared by using the SARPES spectra.

\begin{table}
    \centering
    \caption{The spin-polarization of bands detected in Fig.\ref{fig:Fig2}. ``S", ``B", and ``S(R)" are each denoted for surface, bulk, and surface resonance states; some are bolded to emphasize that the bands A1 and A2 have been already known to originate from the bulk and surface states, respectively, from the previous studies.}
    \begin{threeparttable}
        \setlength{\tabcolsep}{4mm}
        \begin{tabular}{c|c|ccccc}
            \toprule
            \hline
            &Band&1&2&3&4&5  \\
            \hline
            &A&$\downarrow$&$\downarrow$&$\uparrow$&$\downarrow$&$\downarrow$ \\
            ${S_y}$&B&$\downarrow$&&$\downarrow$&$\downarrow$&\\
            \cline{2-7}
            &State&\textbf{S}&\textbf{B}&B&S(R)&\\         
            \hline
            &A&$\uparrow$&&$\downarrow$&$\uparrow$&$\uparrow$ \\
            $S_x$&B&$\uparrow$&&$\uparrow$&$\uparrow$&$\downarrow$ \\
            \cline{2-7}
            &State&\textbf{S}&\textbf{B}&B&S(R)&B\\
            \hline
        \bottomrule    
        \end{tabular}
        \label{Bands}
    \end{threeparttable}
\end{table}

In the upper half of Fig. \ref{fig:Fig2}, we examine the SARPES results of $s_y$ component for the bands of surface A [Figs. \ref{fig:Fig2}(a2)-(a4)] and surface B [Figs. \ref{fig:Fig2}(b2)-(b4)] along cut 3 indicated in Figs. 1(d) and 1(e), respectively; for comparison, high-resolution spin-integrated ARPES images are also shown in Figs. \ref{fig:Fig2}(a1) and (b1). The middle two columns [Figs. \ref{fig:Fig2}(a2,a3), and (b2,b3)] exhibit the SARPES intensity plots of the spin up and down; colored circles indicate the energy states with spin up and down determined from the peak positions of SARPES spectra, extracted in Figs. \ref{fig:Fig2}(a4) and (b4) for several $k$ points (marked as 1 to 4 on the SARPES plots). The band dispersions (dashed lines) determined from the spin-integrated ARPES spectra are duplicated onto the SARPES plots, and instead, the states for spin up and down (plots) are duplicated onto the spin-integrated ARPES plots; here, the same color is used for the bands of surface A and B with the same numbers, such as in A3 and B3. The maps for spin-polarization are shown in Fig. S2 in the Supplemental Material \cite{WTe2PRBSM}.

The correspondence between labeled bands and observed spin polarization is summarized in the upper half of Table \ref{Bands}. Now we use the following two criteria to distinguish surface states from bulk states. 

\begin{itemize}
  \item $\bf {Criterion~1}$: Band A1 is a surface state forming the Fermi arc, whereas band A2 is a bulk state forming the electron pocket. This criterion simply follows the view which has been now commonly acknowledged based on the many previous researches \cite{Wang2016,Wu2016,Bruno2016}. 
  To emphasize it, A1 and A2 are labeled as bold letters ``$\mathbf S$'' (surface) and ``$\mathbf B$'' (bulk), respectively, in table  \ref{Bands}. 
  \item $\bf {Criterion~2}$: The bulk states should exhibit the opposite spin-directions when observing from the opposite surfaces (surface A and B) [see Fig. \ref{fig:Fig1}(b)], and hence, the bands showing the same spin-polarization can be assigned as surface states. This is a key criterion in our study.
\end{itemize}

\begin{figure}[t]
\includegraphics[width=2.5in]{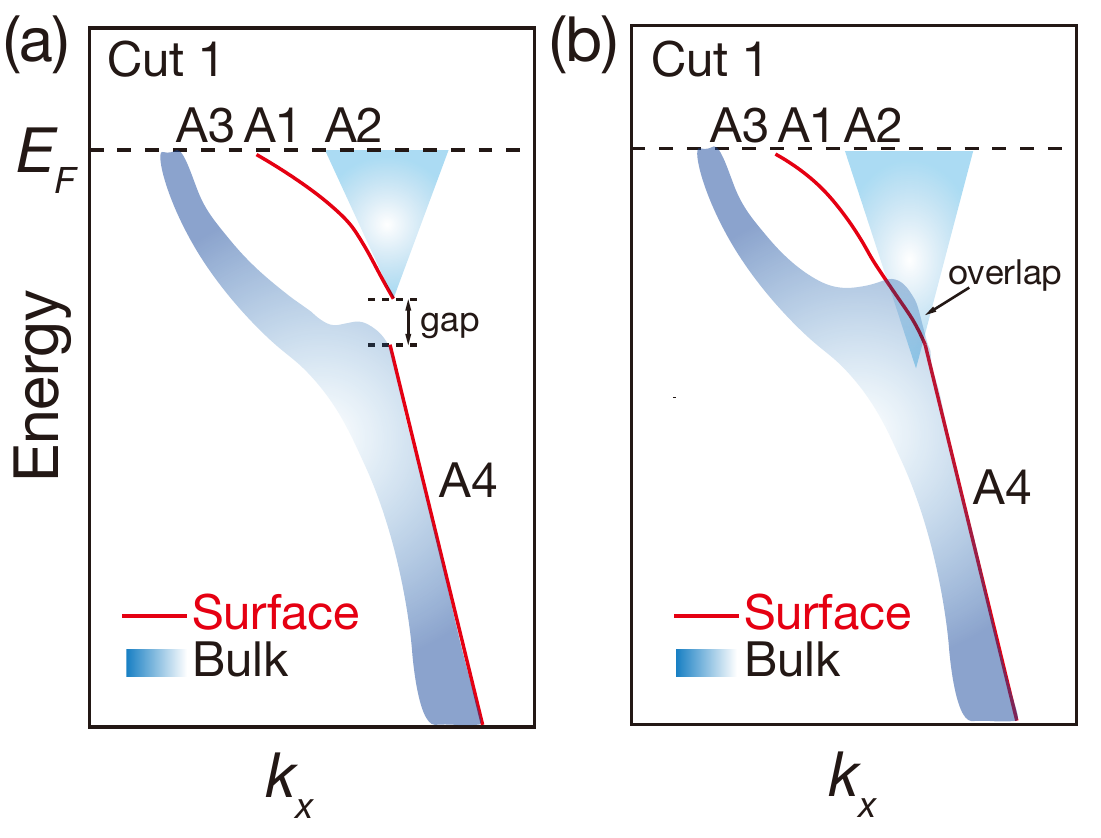}
\caption{Schematics of two different cases for the band structures along cut 1 in Fig. \ref{fig:Fig1}. 
(a) The case predicted by band calculations: A pronounced gap exists between the bulk electron and hole bands. (b) The case we have revealed: The bulk band continuum projected on the surface is overlapped with each other, smoothly connecting the surface resonance band derived from the bulk hole band with the surface band which is from the bulk electron band and forms the Fermi arc.}
\label{fig:Fig3}
\end{figure}

One might think that band B1 should form a pair with band A2 for comparison, rather than with band A1. To examine this, we compared the spin-polarization of band B1 with that of band A2. We found that these have the same spin-direction.  
According to Criterion 2 combined with this result, A2 is assigned as a surface state. This is, however, contradicting with Criterion 1 that A2 should be a bulk state. In contrast, there is no conflict when band B1 and A1, which also show the same spin-direction, are a pair, since Criterion 1 and 2 both expect that band A1 is a surface state. 
Based on our data, we also assign other bands, as described in Table \ref{Bands}, in which the surface state and bulk state are labeled as ``S" and ``B", respectively; some states could not be detected due to the matrix element effect in ARPES, so the labels for these are left as blank in the table. These results are also represented in Fig. \ref{fig:Fig2} by distinguishing the surface and bulk states with open and filled circles, respectively. 

To further confirm our conclusion, we have also performed SARPES measurements along a different cut (cut 2) and a different spin component ($s_x$ component), as shown in the bottom half of Fig. \ref{fig:Fig2}. The determined assignments for the surface and bulk states are summarized in Table \ref{Bands} and distinguished by open and filled circles in Fig. 2, respectively. The results are consistent with those determined from the $s_y$ component, validating our conclusion. 

Our results reveal that the short arc-like feature detected on surface B is a surface state, not a portion of the bulk band. We also found that the bands A4 and B4 originate from surface states, which are most likely for the surface resonance states derived from the bulk hole band; this is compatible with the sharp spectral feature of bands A4 and B4, indicating that these bands are two dimensional, not affected by the $k_z$ broadening. Note that surface resonance bands dispersing nearby the bulk band projection commonly appear together with topological surface state \cite{hsieh2010direct,yan2015topological,jozwiak2016spin,Hasan}. We emphasize here that our conclusions became available by our novel method of distinguishing the bulk and surface states via a comparison of the SARPES spectra between the (001) and ($00\overline{1}$)  crystal surfaces.

\begin{figure*}
\includegraphics[width=5in]{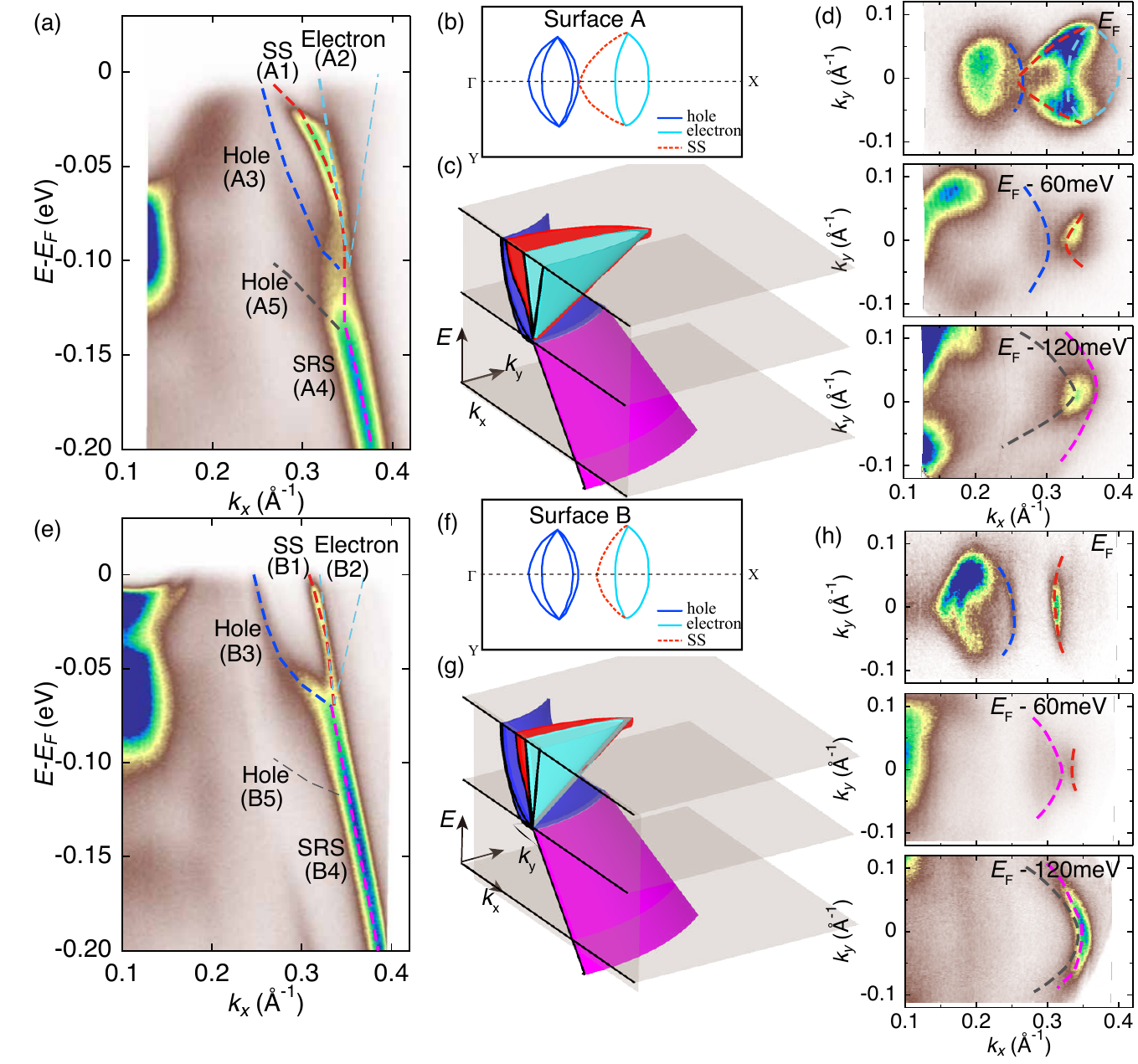}
\caption{(a) The assignments (colored dashed lines) for each segment of the band we observed along the high-symmetry cut 1 on surface A, overlapped on the ARPES image [the same image as that in Fig. \ref{fig:Fig1}(f)]: red for surface state forming Fermi arc, light-blue for bulk electron band, blue for bulk hole band, and purple for surface resonance state, which are the same colors as in Fig. \ref{fig:Fig2}. (b,c) Schematic Fermi surface and band structures on surface A, illustrated with the same colors as in (a). The transparent $k_x-k_y$ planes are drawn at three different energies, more or less corresponding to those in the ARPES plots plotted in (d). (d) ARPES plots along $k_x-k_y$ obtained at three different energies ($E_{\rm F}$, $E_{\rm F}-$60 meV, and $E_{\rm F}-$120 meV) on surface A. The colored dashed lines correspond to the bands assigned in (a). (e)-(h) The ARPES data and schematic band structures similar to those in (a-d), but for surface B.}
\label{fig:Fig4}
\end{figure*}

Interestingly, the band A4 and B4 (surface resonance states) are connected with the bands A1 and B1 (surface states forming Fermi arc) along the high symmetry cut 1 [see Fig. \ref{fig:Fig1}(f) and Fig. \ref{fig:Fig1}(g)], respectively. This behavior is schematically illustrated in Fig. \ref{fig:Fig3}(b). The intensities around the connection between A1 and A4 seem to be weak [Fig. \ref{fig:Fig1}(f)] due to the matrix element effect. A smooth connection is confirmed more clearly in the data with $s$ polarized light, as shown in Fig. S1 of the Supplemental Material \cite{WTe2PRBSM}. In surface band calculations of WTe$_2$ \cite{Soluyanov2015,Bruno2016,Wang2016}, a pronounced gap is opened between the electron and hole bands projected onto the surface, as illustrated in Fig. \ref{fig:Fig3}(a); nevertheless, this is not consistent with our data. Previously, it was pointed out that the distance of electron and hole bands are much closer in momentum space and the discrepancy from measurements can be mitigated by taking only several layers into account in the theoretical calculations \cite{Das2016}.  Our new data indicates that the electron and hole bands are very close not only in momentum but also in energy, thus those projected on the surface are overlapped with each other; consequently, the dispersions of surface states derived from the bulk electron and hole bands are mutually connected.

In Figs. \ref{fig:Fig4}(a) and (e), we plot the ARPES image along $\Gamma-$X for surfaces A and B, respectively, and summarize our results by describing the character of each band determined with the SARPES spectra: Fermi-arc surface state (SS), bulk hole band (Hole), bulk electron band (Electron), and surface resonance state (SRS), which are drawn by dashed curves with same colors for the bands corresponding between surface A and B. These Fermi surfaces and band structures are schematically illustrated in Figs. \ref{fig:Fig4}(b,c) and Figs. \ref{fig:Fig4}(f,g) for surface A and B, respectively. Our data revealed that the surface Fermi-arc states exist on both surfaces A and  B [or (001) and ($00\overline{1}$)  crystal surfaces of WTe$_2$]. The surface resonance bands (magenta color) merge into the hole band (blue color), and are also smoothly connected with the Fermi-arc surface bands (red color). 

The band shapes are further confirmed in Figs. \ref{fig:Fig4}(d) and (h) by plotting the ARPES plots along $k_x-k_y$ sheets at different energies. On surface A [Fig. \ref{fig:Fig4}(d)], the surface state forms a large Fermi arc at $E_F$ (red dashed line in the top panel) extending from the electron pocket up to the hole pocket. With increasing binding energy, the arc gets smaller (the middle panel), and it merges with the surface resonance band, which forms an arc structure with the opposite curvature and is enlarged with the further increase of the binding energy (magenta dashed line in the bottom panel). On the other hand, the Fermi arc on surface B is much smaller [the upper panel of Fig. \ref{fig:Fig4}(h)], being closely adjacent with the bulk electron pocket (not clearly visible due to the matrix element effects) without extending toward the hole band. The evolution of band structures with binding energy on surface B is similar to that on surface A: the arc gets smaller and merges with the surface resonance band, which forms an arc shape with the opposite curvature (the middle and bottom panels). 
Our results, thus, reveal intriguing band dispersions originated from the surface states for both crystal surfaces, which have a Dirac-cone-like shape, which is reported in previous studies \cite{Wu2015,Wu2016,Thirupathaiah2017,Belopolski2016}, but half cut on opposite sides in the upper and lower cones, as illustrated by red and magenta dispersions in Figs. \ref{fig:Fig4}(c) and (g). 

In summary, we prepared pairs of samples with the (001) and ($00\overline{1}$)  surfaces (surface A and B) of WTe$_2$ by ``sandwich method", and separately observed band structure of each surface by high-resolution laser-ARPES and SARPES. The surface states were distinguished from the bulk states by comparing the SARPES spectra of surfaces A and B, which revealed the presence of Fermi arcs with the surface state origin on both the top and bottom surfaces. We found a surface resonance band adjacent to the bulk hole band, which forms arc structures at energies below $\sim$ -100 meV. Interestingly, the surface resonance band smoothly connects with the band which forms the Fermi arc, yielding a Dirac-like band, which is half cut in opposite sides of the upper and lower cones. Our results indicate that the bulk electron and hole bands are so close in momentum space that their continuum projected onto the surfaces are overlapped with each other, disagreeing with band calculations. Although the present results alone cannot be decisive to conclude that WTe$_2$ is indeed a Weyl semimetal~\cite{Bruno2016}, these will provide severe constraints in a complete understanding of the topological nature of  this compound. Furthermore, we note that a theoretical research on bcc iron, which has Weyl points, indicates that there are topological nontrivial surface resonance states connecting with the topological Fermi arcs \cite{gosalbez2020topological}. It suggests that the surface resonance state we found in WTe$_2$ may also have a deeper physical meaning and would inspire the follow-up researches.


This work was supported by the JSPS KAKENHI (grant numbers JP18H01165, JP18K03484, JP19H02683, JP19F19030 and JP19H00651), MEXT Q-LEAP (grant number JPMXS0118068681), and by MEXT under the “Program for Promoting Researches on the Supercomputer Fugaku” (Basic Science for Emergence and Functionality in Quantum Matter Innovative Strongly Correlated Electron Science by Integration of “Fugaku” and Frontier Experiments) (Project ID: hp200132).
The work at Rutgers University was supported by the NSF under Grant No. DMR-1629059, and that at Postech was supported by the National Research Foundation of Korea (NRF) funded by the Ministry of Science and ICT (No. 2016K1A4A4A01922028).


%

\end{document}